\documentclass[11pt,twocolumn]{article}
\usepackage[affil-it]{authblk}
\usepackage{fullpage,amsmath,amssymb,graphicx,verbatim,color,hyperref}
\usepackage[top=1in,bottom=1in,left=0.5in,right=0.5in]{geometry}
\usepackage{cite}

\usepackage{graphicx} 
\usepackage{lastpage}
\usepackage[format=plain,justification=raggedright,singlelinecheck=false,font=small,labelfont=bf,labelsep=space]{caption} 
\usepackage{tabularx}
\usepackage{abstract}

\usepackage{amsmath,amssymb}
\usepackage{tabularx}
\usepackage{array}
\usepackage{upgreek}
\usepackage[euler]{textgreek}
\newcommand{\microns}{\,\upmu\text{m}}

\newcommand{\be}{\begin{equation}}
\newcommand{\ee}{\end{equation}}
\newcommand{\bea}{\begin{eqnarray}}
\newcommand{\eea}{\end{eqnarray}}
\newcommand{\bex}{\begin{equation*}}
\newcommand{\eex}{\end{equation*}}
\newcommand{\beax}{\begin{eqnarray*}}
\newcommand{\eeax}{\end{eqnarray*}}

\newcommand{\unit}{\;\mathrm}

\begin{document}
\renewcommand*{\thefootnote}{\fnsymbol{footnote}}

\title{Shape-controlled orientation and assembly of colloids with sharp edges in nematic liquid crystals}

\author[1]{Daniel A.\ Beller\thanks{Electronic address: dbeller8@gmail.com}}
\author[1,2]{Mohamed A.\ Gharbi}
\author[2]{Iris B.\ Liu}

\affil[1]{Department of Physics and Astronomy, University of Pennsylvania, Philadelphia, Pennsylvania 19104, USA}
\affil[2]{Department of Chemical and Biomolecular Engineering, University of Pennsylvania, Philadelphia, Pennsylvania 19104, USA}

\date{}

\twocolumn[
\maketitle
\begin{onecolabstract}The assembly of colloids in nematic liquid crystals via topological defects has been extensively studied for spherical particles, and investigations of other colloid shapes have revealed a wide array of new assembly behaviors. We show, using Landau-de Gennes numerical modeling, that nematic defect configurations and colloidal assembly can be strongly influenced by fine details of colloid shape, in particular the presence of sharp edges. For cylinder, microbullet, and cube colloid geometries, we obtain the particles' equilibrium alignment directions and effective pair interaction potentials as a function of simple shape parameters. We find that defects pin at sharp edges, and that the colloid consequently orients at an oblique angle relative to the far-field nematic director that depends on the colloid's shape. This shape-dependent alignment, which we confirm in experimental measurements, raises the possibility of selecting self-assembly outcomes for colloids in liquid crystals by tuning particle geometry.
\end{onecolabstract}
\vspace{0.5cm}
]
\saythanks







\section{Introduction}

A promising route toward targeted colloidal self-assembly is to disperse colloids in a nematic liquid crystal, whose elasticity resists the deformations caused by anchoring conditions at the surfaces of the colloids, resulting in forces on the colloids that push them into ordered arrangements such as chains and triangular lattices \cite{poulin1997novel, Musevic2006, smalyukh2004ordered, Gharbi2011anchoring, Skarabot2008}. Nematic colloidal dispersions are especially interesting because each colloidal particle is accompanied by a topological defect or defects in the liquid crystal that play a key role in the assembly process \cite{ terentjev1995disclination,  poulin1997novel,Musevic2006}. Spherical colloids have been extensively studied in this context. When the spherical colloid's surface imposes strong homeotropic (normal) anchoring conditions on the nematic director field, the companion defect is either a point-like hedgehog or a linear ``Saturn ring'' disclination about the equator. Ring disclinations can be manipulated, such as by laser tweezers, to topologically entangle multiple colloids into chains\cite{Ravnik2007} and knotted structures \cite{Tkalec:2011lj}. Because the colloidal lengthscale roughly coincides with the wavelengths of visible light, colloidal structures assembled by nematic elasticity and topological defects hold strong potential for photonics applications such as photonic bandgap crystals \cite{nych2013assembly,muvsevivc2011direct, ravnik2011three}.

While much of the research on nematic colloidal dispersions has concentrated on spherical colloidal particles, colloidal self-assembly in general is known to depend strongly on the shapes of the colloidal particles. This has been amply demonstrated in the entropically-driven self-assembly of hard particles \cite{damasceno2012predictive, 
C2SM25813G, 
de2011dense} and in the capillarity-driven assembly of colloids at fluid interfaces \cite{botto2012capillary}. Recent advances in colloid fabrication techniques, such as photolithography, have enabled fairly precise creation of particles with designed asymmetric shapes as well as surface functionalization \cite{Lee2011195}.   In nematics, a number of recent investigations have begun to explore how defect configuration and assembly behavior change when the particle has a non-spherical geometry, such as rods \cite{tkalec2008interactions}, spherocylinders\cite{HungPRE2009}, ellipsoids \cite{tasinkevych2014dispersions}, microbullets (rods with one flat end and one rounded end) \cite{gharbi2013microbulletPublished}, polygonal platelets \cite{lapointe2009shape, dontabhaktuni2012shape}, washers\cite{cavallaro2013ringSoftMatter}, handle-bodies of varying genus \cite{senyuk2012topological}, and M\"obius bands \cite{machon2013knots}.  In addition, nontrivial colloid or substrate geometry has been shown to provide elastically attractive or repulsive sites influencing the organization of colloids in the bulk or at an interface \cite{cavallaro2013ringSoftMatter, cavallaro2013exploitingPNAS, PhysRevLett.112.225501, 2014arXiv1406.0702E}.

While these investigations have found a variety of phenomena resulting from gross changes to the colloid shape, or even topology, as compared to spheres, it is important to also keep in mind that fine details of colloid shape can play a surprisingly important role. In particular, sharp edges often act as pinning sites for nematic defects because of the elastic strain that they impose through a large variation in the anchoring direction over a small region \cite{tsakonas2007multistable, luo2012multistability, cavallaro2013exploitingPNAS}. (An abrupt change in anchoring chemistry from degenerate planar to homeotropic has a similar pinning effect \cite{conradi2009janus, vcopar2014janus}.) Therefore, a rod-like colloid with rounded ends, such as a spherocylinder or ellipse, may behave quite differently from a cylinder. These distinctions prompt caution but also suggest a path to tuning colloidal assembly properties by varying the details of colloid shape.

In this paper, we investigate the topological defect configurations and assembly properties of colloids with sharp edges, in the shapes of cylinders, microbullets, and cubes. We focus primarily on the case of colloids with strong homeotropic anchoring confined in thin planar cells, in the regime where the companion topological defect is a disclination ring (valid for colloids of small size\cite{stark1999director} and/or under strong confinement\cite{grollau2003spherical}). Through Landau-de Gennes numerical modeling, we show that the edges of these shapes result in significantly different defect ring configurations than those seen with smooth shapes such as spheres and ellipsoids. The quadrupolar symmetry of the smooth rods' Saturn ring configuration is lost for shapes with edges; instead, defect rings pin at portions of the edges. As a result, cylinders and microbullets orient at oblique angles relative to the far-field nematic director, an effect not seen at equilibrium in smooth rods such as spherocylinders\cite{HungPRE2009} or ellipsoids \cite{tasinkevych2014dispersions}. This oblique angle increases  with the aspect ratio of the colloid, as we confirm with experimental observations, demonstrating tunable colloidal alignment and interactions through particle geometry. 

In addition, the disclination configuration spontaneously breaks a reflection symmetry or symmetries, implying multistable colloidal orientations. We find that this multistability strongly affects not only colloidal pair interactions but also the defect behavior on individual colloids, which may exhibit sudden jumps between defect configurations in response to an applied torque. 
Our results highlight the importance of details of particle geometry in relation to topological defects and colloidal interactions, and suggest new routes toward complex and multi-stable self-assembled structures in nematic colloidal dispersions.

\section{Numerical approach}

We study the nematic liquid crystal using Landau-de~Gennes (LdG) numerical modeling \cite{Ravnik2009a}. The nematic state is represented as a traceless symmetric tensor field $\mathbf{Q}(\vec{x})$, which in a uniaxial nematic with director $\vec{n}$ satisfies $Q_{ij}=\tfrac{3}{2} S(n_i n_j -\tfrac{1}{3} \delta_{ij})$. The leading eigenvalue $S$ is the nematic degree of order, and we visualize defects as isosurfaces where $S$ falls to $3/4$ of its bulk value $S_0$ in an undistorted nematic, except where otherwise indicated. The LdG free energy in the single-elastic constant limit, 
\begin{align}
F_{\mathrm{LdG}} &= \int dV  \left[\tfrac{1}{2} A \mathrm{tr}\left(\mathbf{Q}^2\right) + \tfrac{1}{3} B \mathrm{tr}\left(\mathbf{Q}^3\right)+\tfrac{1}{4} C \left(\mathrm{tr}\left(\mathbf{Q}^2\right)\right)^2 \right. \nonumber \\
& \qquad \qquad \left. + \frac{L}{2} \frac{\partial Q_{ij}}{\partial x_k}\frac{\partial Q_{ij}}{\partial x_k}\right], \label{FLdG}
\end{align}
is minimized numerically in a finite difference scheme on a regular cubic mesh with grid spacing 4.5 nm, using a conjugate gradient algorithm from the ALGLIB package (www.alglib.net). We take from Ref.~\cite{Ravnik2009a} the typical values for the material constants of the commonly used nematic liquid crystal 5CB: $A=-0.172\times10^6\;\mathrm{J/m^3}$, $B=-2.12\times10^6\;\mathrm{J/m^3}$, $C=1.73\times10^6\;\mathrm{J/m^3}$. 
We set the elastic constant $L=8\times10^{-12}\;\mathrm{N}$ to correspond to a Frank elastic constant $K=L\cdot 9S_0^2/2\approx 1\times 10^{-11}\unit{N}$ roughly matching the elastic constants of 5CB \cite{gharbi2013microparticles}. 
We do not include a surface energy term because we study the limit of infinitely strong anchoring, where $\mathbf{Q}$ is held fixed at the boundaries to give a fixed boundary director \cite{hung2009faceted}. (We have checked that replacing this assumption with the presence of a Nobili-Durand surface anchoring term $W \int dA \mathrm{tr}\left(\left(\mathbf{Q}-\mathbf{Q}^s\right)^2\right)$, where $\mathbf{Q}^s$ is the $\mathbf{Q}$ favored by the surface, in $F_{\text{LdG}}$ does not change the results significantly for experimentally realistic anchoring strength $W$ in the strong anchoring regime.)

The top and bottom boundaries of the rectangular cell have oriented planar anchoring along a uniform direction, the ``rubbing direction'', which defines the bulk far-field director $\vec{n}_0$. At the sides of the box, we use free boundary conditions, eschewing periodic boundary conditions to ensure that a colloid does not interact with its periodic images. The colloid, with infinitely strong homeotropic anchoring, is placed at the mid-height of the cell, and its orientation is held fixed while the nematic configuration $\mathbf{Q}(\vec{x})$ is relaxed. The initial condition consists of a uniform uniaxial nematic with director $\vec{n}=\vec{n}_0$ everywhere except the boundaries; we have checked in several instances that random initial conditions produce either the same states or metastable states of higher energy. 

When studying the effects of colloid shapes with edges, we have to take care not to make the modeled edges {\em too} sharp because perfectly sharp edges introduce an unrealistic ambiguity into the director field: The rotation of the director through $\pm90^\circ$ can be accomplished through either splay or bend, which have equal energy in the one-elastic-constant approximation. Because the colloidal edges are expected to be rounded in a convex way at least at the molecular scale, 
the splay configuration is more realistic. In order to obtain this configuration, we round the edges of numerically modeled cylindrical colloids using the ``superegg'' equation,
\begin{equation} \left(\frac{x^2+y^2}{b^2}\right)^p+\left(\frac{z^2}{a^2}\right)^p = 1. \label{supereggeqn} \end{equation}
Here the coordinate frame has been rotated so that the colloid long axis points along the $z$ direction; $b$ is the colloid radius; $a$ is half the colloid length; and $p$ is a parameter we refer to as the ``sharpness'' of the shape, interpolating between an ellipsoid at $p=1$ and a right circular cylinder as $p\rightarrow \infty$. We treat $p$ as an adjustable shape parameter controlling the degree to which curvature is concentrated in edges. To model microbullets, only the half of the colloid opposite the hemispherical cap is given the superegg shape; the other half is spherocylindrical, with no edges. 

Similarly, to model colloidal cubes, we round the edges using a special case of the superellipsoid equation, \begin{equation}x^{2p} + y^{2p} + z^{2p} = a^{2p}.\label{superellipsoideqn}\end{equation} Here,  $p$ interpolates the shape between a sphere of radius $a$ at $p=1$ and a cube of side length $2a$ as $p\rightarrow\infty$. (For simplicity, we do not vary the aspect ratio for this shape.)

\section{Colloidal cylinders and microbullets: Shape-controlled orientation}

Before presenting our results, we recall the known behavior of homeotropic spherical particles in a nematic. 
A spherical colloid imposing strong homeotropic anchoring on the director field at the colloid's surface acts as a radial hedgehog, and is thus accompanied by a stable topological defect to ensure zero total topological charge \cite{mermin1979topological}. This companion defect can be either a hyperbolic point hedgehog or a ``Saturn ring'' disclination of unit topological charge \cite{terentjev1995disclination}. In the latter case, the ring's stable position is around the equator of the colloid, in the plane orthogonal to the far-field nematic director $\vec{n}_0$, and the director field has quadrupolar symmetry \cite{Lubensky1998}.  For an isolated spherical colloid, the director field in a cross-section transverse to the disclination tangent is that of a 2D nematic disclination with winding number $-1/2$. 
 In the case of a hedgehog defect, the director field has dipolar symmetry, with the hedgehog separated from the colloid center along the $\vec{n}_0$ direction. Smaller colloid size\cite{stark1999director} or stronger confinement\cite{grollau2003spherical} tends to stabilize the Saturn ring relative to the hedgehog. These sphere-defect pairs self-assemble into chains parallel to $\vec{n}_0$ in the case of hedgehog defects, and kinked chains making an angle $\sim 73^\circ$ with $\vec{n}_0$ in the case of Saturn rings \cite{Musevic2006, Skarabot2008}. Saturn rings of neighboring colloids may also join into a single disclination ring that entangles pairs or chains of colloids \cite{Ravnik2007}.

For homeotropic colloids that are topologically but not geometrically spheres, such as cubes and cylinders (with edges at least slightly rounded), the same topology applies to the surrounding nematic: Each genus-zero colloid is accompanied by a companion defect, either a hedgehog or a disclination ring \cite{tkalec2008interactions, hung2009faceted, HungPRE2009}. Colloidal micro-rods have been shown to form chains and bound pairs depending on the types and positions of their defects, which were either a hedgehog near the micro-rod's end or a disclination ring wrapping around the colloid's long axis \cite{tkalec2008interactions}. 

Our LdG numerical modeling of rod-like colloids shows that the details of particle shape strongly affect both the defect configuration and the colloids' alignment and assembly. We numerically model homeotropic ellipsoids, cylinders, and microbullets confined in thin cells of thickness three times the colloid diameter, with oriented planar anchoring at the boundaries, and we seek the equilibrium angle $\phi_0$ of the rod-like colloid's long axis (assumed to lie in the horizontal plane) relative to the far-field director $\vec{n}_0$. 

\begin{figure}
\begin{center}
\includegraphics[width=\linewidth]{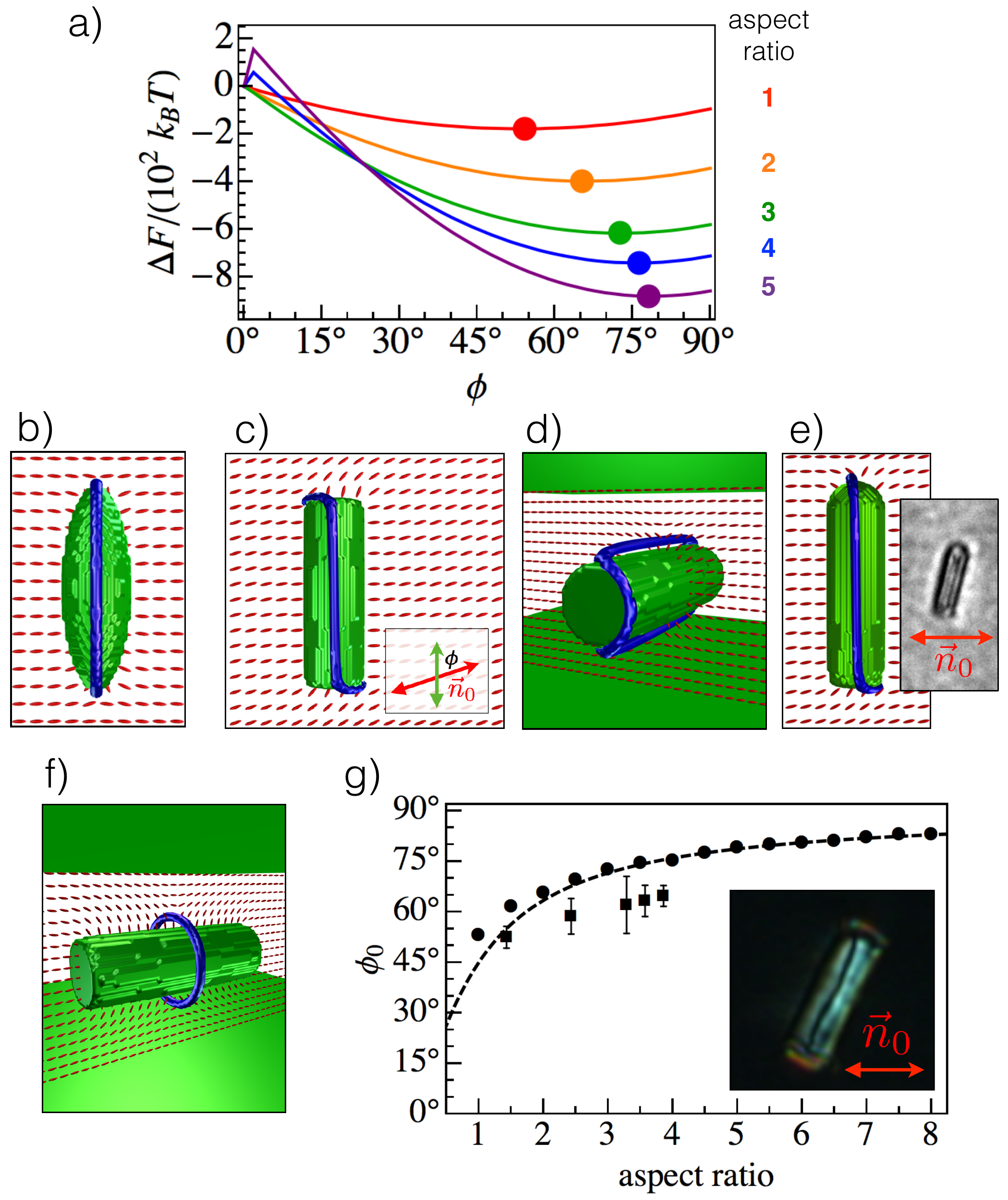}
\caption{Rod-like colloids. (a)  Free energy of a colloidal cylinder (diameter 90 nm), modeled as a $p=10$ superegg, in a planar nematic cell (thickness 270 nm) with strong unidirectional planar anchoring. Free energy is plotted  as a function of the angle $\phi$ between the cylinder's long axis and the far-field director $\vec{n}_0$, with $\phi=0$ taken as the $\Delta F=0$ reference state for each aspect ratio. 
(b-e) Equilibrium disclination configuration around an ellipsoid (b), a cylinder (c,d), and a microbullet (e), with defects in blue and the director field in red; all have diameter 90 nm, aspect ratio 3, and sharpness $p=10$. Inset to (e) shows an experimental bright-field optical image of a microbullet of diameter $2\microns$ and length $10\microns$ at is equilibrium tilt.  (f) An alternative disclination arrangement for cylinders found numerically that is responsible for the sharp peaks in (a) at small $\phi$ for aspect ratios 4 and 5. (g) Comparison of the energy-minimizing cylinder axis angle $\phi_0$ predicted numerically (circles) with experimental data (squares) and the inverse tangent of the aspect ratio (dashed curve). Inset: Experimental polarized optical microscopy image of a colloidal cylinder of length $17\microns$ at its equilibrium oblique tilt $\phi_0$.}
\label{rodsfig}
\end{center}
\end{figure}

For ellipsoids, we find that the disclination ring wraps around the colloid's long axis, as shown in Fig.\ \ref{rodsfig}b, and that the colloid prefers to orient at $\phi=90^\circ$, with the long axis perpendicular to $\vec{n}_0$ \cite{tasinkevych2014dispersions}. The results are markedly different in the case of cylinders.  Figures \ref{rodsfig}c,d show the defect configuration around a cylinder  modeled as a superegg using Eqn.~\ref{supereggeqn}, with sharpness $p=10$.  The disclination ring still wraps around the long axis of the colloid, but something new happens at the sharp edges at the ends of the cylinder: The disclination loop follows an ``S''-shaped contour, as viewed from above, with sharp turns to follow one half of the circular edge at one end and the opposite half of the circular edge at other end. This defect configuration represents an energetic compromise between the configuration preferred by the ellipsoid and the preference for defects to pin at sharp edges to reduce splay.

\begin{figure}
\begin{center}
\includegraphics[width=\linewidth]{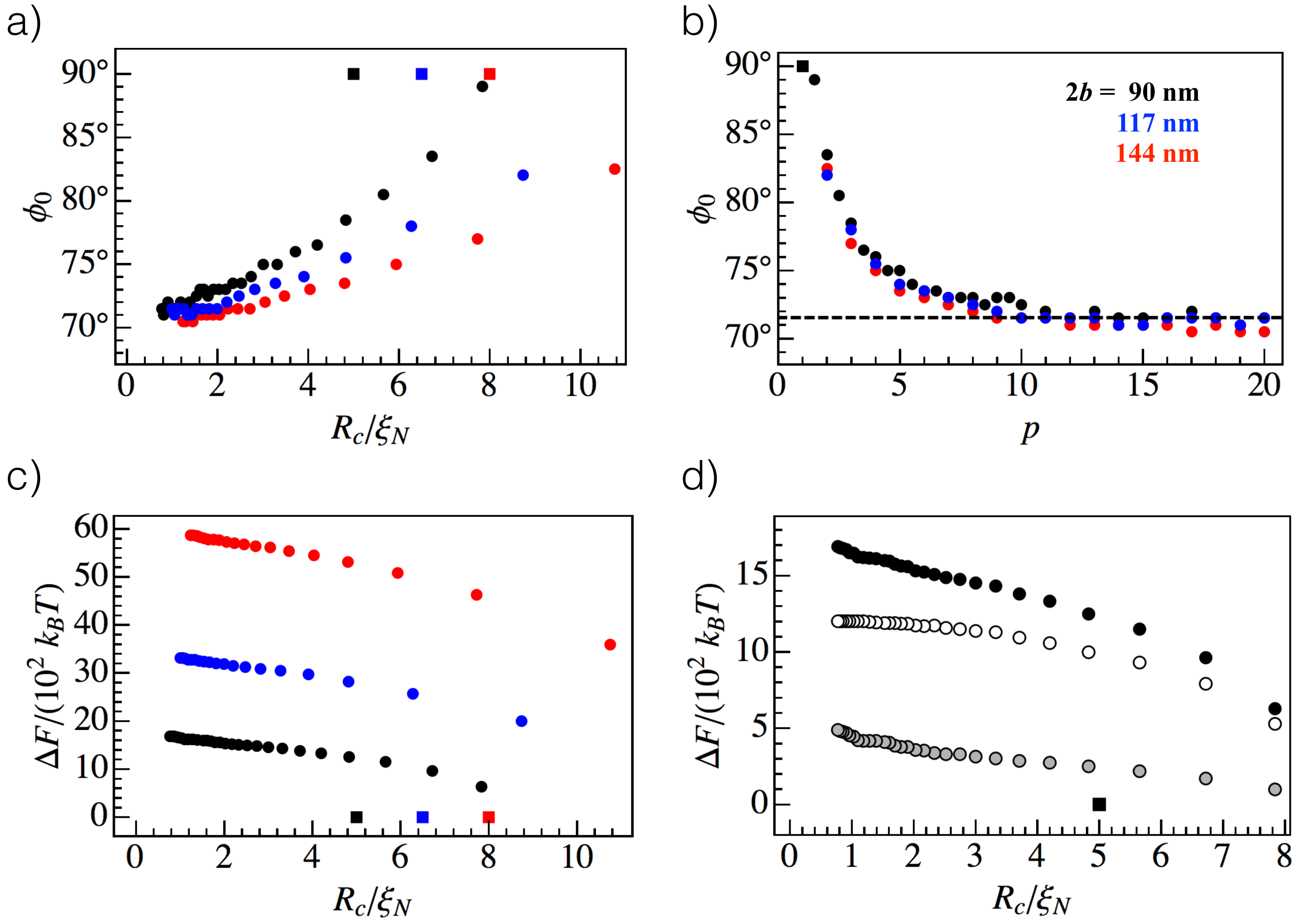}
\caption{Effect of colloid edge sharpness: Data for cylinders, approximated as supereggs, of aspect ratio $a/b=3$ and diameter $2b=90$ nm (black), 117 nm (blue), and 144 nm (red). (a) Equilibrium angle $\phi_0$ versus the radius of curvature $R_c$ of the edge (minimum radius of curvature of the superellipse cross-section), in units of the nematic correlation length $\xi_N\approx 3.0$ nm. Square symbols show data for $p=1$ ellipsoids, where for comparison $R_c$ is taken to be the radius of curvature at the ends. (b) Equilibrium angle $\phi_0$ plotted against sharpness parameter $p$. Dashed line shows analytic approximation $\phi_0=\tan^{-1}(a/b)$. (c) Free energy compared to that of $p=1$ ellipsoid (square symbols). (d) For the superegg with diameter 90 nm, the total free energy (black disks), and its bulk (white disks) and elastic (gray disks) free energy components, compared to that of $p=1$ ellipsoid (black square).   }
\label{figsharp}
\end{center}
\end{figure}

 As a result of this new defect configuration, the cylinder prefers to orient with its long axis at an oblique angle $\phi_0$ relative to $\vec{n}_0$, and $\phi_0$ increases toward $90^\circ$ with increasing  aspect ratio $a/b$. Figure \ref{rodsfig}a shows the change in free energy as cylinders of various aspect ratios are rotated by an angle $\phi$ relative to $\vec{n}_0$.  For aspect ratios $a/b\gtrsim 2$, $\phi_0$ is well approximated by $\tan^{-1}(a/b)$, as shown in Fig.~\ref{rodsfig}g, suggesting that (at least heuristically) the energy is minimized when the separation vector between the kinked ends of the disclination lies in the plane perpendicular to $\vec{n}_0$. 
This analytic approximation improves in accuracy as the sharpness $p$ is increased, as shown in Fig.~\ref{figsharp}(b), while the qualitative trend of $\phi_0$ increasing with aspect ratio persists with decreasing $p$ down to $p\approx 2$. 
For oblate cylinder shapes with aspect ratios below $1$ (down to $a/b=0.25$, the lowest tested), the kinked disclination configuration and oblique colloid orientation remain, but the positive correlation of $\phi_0$ with $a/b$ disappears.

An alternative configuration, with the disclination ring encircling the colloid's short axis about its center (Fig.~\ref{rodsfig}f), is also observed at high aspect ratio, but this state only appears when the colloid is constrained to orient at unfavorably small angles $\phi$ relative to $\vec{n}_0$. (Sometimes the ring is localized at one of the cylinder ends instead of about its center.) 
If the aspect ratio is increased sufficiently, we expect that the defect's core  energy per unit length might allow this short-axis disclination ring to eventually overtake the kinked long-axis disclination loop in stability, resulting in alignment of long cylindrical colloids parallel to the far-field director at $\phi=0$. However, in the numerical results, the kinked long-axis disclination loop remains stable, with $\phi_0\approx \tan^{-1}(a/b)$, for aspect ratios up to $a/b=10$, the highest tested. 

We find a similar result for microbullets (Fig.~\ref{rodsfig}e). Here, the disclination again wraps around the colloid's long axis and turns sharply to follow half of the circular end. At the rounded end, the disclination is nearly undeflected. The energetically preferred colloidal orientation $\phi_0$ is again oblique and increases with increasing aspect ratio. At all aspect ratios tested, $\phi_0$ is well approximated simply by the average of $\tan^{-1}(a/b)$ and $90^\circ$. 

Experimental measurements on colloidal cylinders in the nematic liquid crystal 5CB are in qualitative agreement with these numerical predictions. Colloidal cylinders of diameter $7\microns$ and varying aspect ratios are fabricated using standard lithographic techniques with SU-8, a negative photoresist. To impose homeotropic anchoring of the nematic at their surfaces, the cylinders are covered with a thin layer of chromium (around 30 nm) then treated with a 3\%wt solution of N, N-dimethyl-N-octadecyl-3-aminopropyltrimethoxysilyl chloride (DMOAP) in a mixture of 90\%wt ethanol and 10\%wt water. Two glass coverslips, treated with  a 1\% wt solution of polyvinyl alcohol (PVA) in a mixture of 5\%wt ethanol and 95\%wt water and rubbed along one direction after heating, are used to create a thin planar cell of thickness approximately $10\microns$ (fixed using Mylar spacers). A 1\%wt dispersion of colloidal cylinders in 5CB is introduced into the LC cells via capillarity. 

We observe that cylinders with hedgehog companion defects align parallel to $\vec{n}_0$, while many cylinders with disclination rings are wrapped by their defects in an approximately straight line along the cylinder axis as viewed from above, consistent with the ``S''-shaped contour profile predicted numerically (see inset to Fig.~\ref{rodsfig}g). For the latter configuration, the cylinders are oriented at an oblique angle $\phi_0$ between $45^\circ$ and $90^\circ$ that increases with the cylinder's aspect ratio (Fig.~\ref{rodsfig}g). We observe a similar oblique $\phi_0$ for microbullet colloids with ring defects (see inset to Fig.~\ref{rodsfig}e), fabricated as described in Ref.~\cite{gharbi2013microbulletPublished}. We also note that, in Ref.~\cite{tkalec2008interactions}, a wide range of micro-rod orientations $35^\circ \lesssim \phi_0 \lesssim 90^\circ$ was reported for a wide range of aspect ratios $1.3 \lesssim a/b \lesssim 4.7$.

\begin{figure}
\begin{center}
\includegraphics[width=\linewidth]{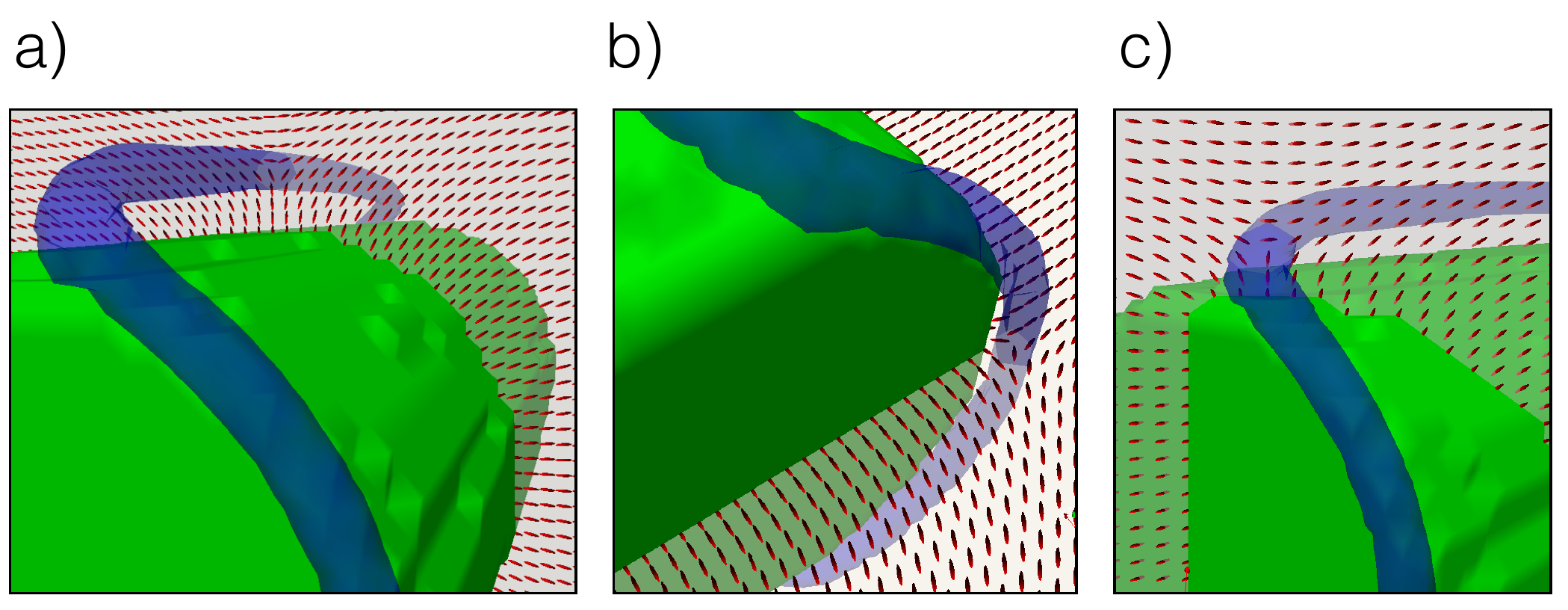}
\caption{Detailed view of the director field around a cylindrical colloid. The cylinder in this case is a $p=10$ superegg with diameter $2b=144$ nm and aspect ratio $a/b=3$. The spacing between rods in the director field profile is 4.5 nm, or 1.5 times the nematic correlation length. (a) Director profile in a plane normal to the colloid's long axis, near the colloid's center. (b) Director profile in a plane parallel to the planar boundaries, at one end of the cylinder. (c) Director profile in a plane oblique to the colloid, passing through a kink in the disclination. }
\label{figcloseup}
\end{center}
\end{figure}

Detailed views of the numerically observed director field profile about the disclinations are given in Fig.~\ref{figcloseup} for a cylindrical colloid approximated as a superegg of diameter 144 nm, aspect ratio 3, and sharpness $p=10$. There it is seen that the director profile in a plane perpendicular to the defect tangent is always approximately that of a planar  nematic disclination with winding number $-1/2$. This is true on both  the approximately straight and approximately semicircular segments of the disclination, as well as the disclination kinks where these different segments meet. On the other hand, the disclination's distance from the colloid surface changes significantly between the different segments. On the approximately straight segments that travel along the cylinder's long axis, the disclination is roughly 18 nm, or about 6 times the nematic correlation length $\xi_N\approx 3.0$ nm, away from the  colloid surface, whereas along the semicircular segments along the colloid edges, the distance is about half as great. For spherical colloids, it has previously been found numerically that the distance of the Saturn ring from the colloid surface increases with colloid radius \cite{stark1999director}.   (Here we refer to absolute distance, not its ratio to the colloid radius). The variation of defect-colloidal surface separation around a single cylindrical colloid in Fig.~\ref{figcloseup} suggests that this principle holds for local curvature, as well.

\begin{figure}
\begin{center}
\includegraphics[width=\linewidth]{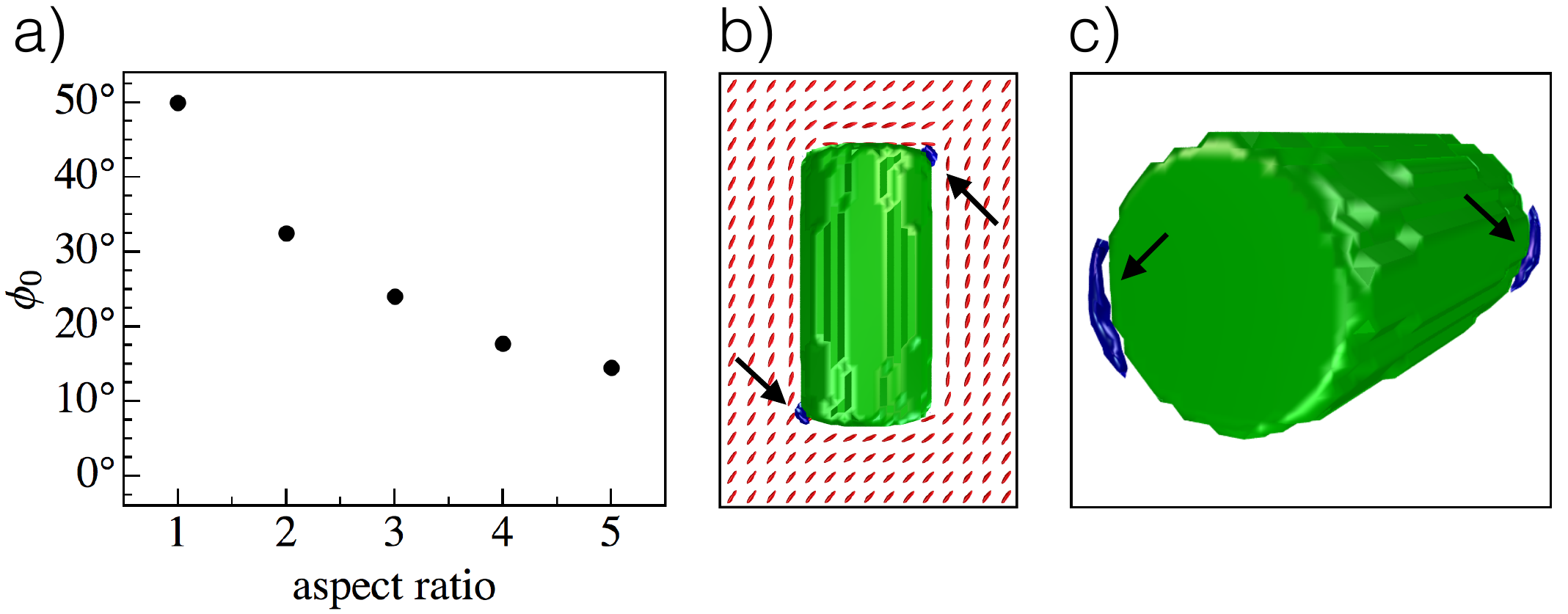}
\caption{Equilibrium orientation of cylindrical colloids with strong degenerate planar anchoring. The colloids have diameter $2b=90$~nm and sharpness parameter $p=10$. (a) Dependence of equilibrium colloid orientation $\phi_0$ on the aspect ratio. (b) Director field and boojum defect structure for a cylinder with aspect ratio 2, viewed from above. For better visibility, the boojums are visualized here as isosurfaces of $S=0.85 \times S_0$ and indicated with arrows. (c) The same cylinder and its defects viewed from the side, showing that the regions of melted nematic at the boojums (again indicated with arrows) extend along portions of the edges.}
\label{figplanar}
\end{center}
\end{figure}

We also observe a dependence of colloid orientation $\phi_0$ on aspect ratio when the colloid's anchoring conditions are degenerate planar rather than homeotropic. However, the change in $\phi_0$ reverses, now decreasing with increasing aspect ratio (Fig.~\ref{figplanar}a). In other words, as the aspect ratio increases the colloid long axis orients increasingly toward $\vec n_0$ rather than toward $\hat z \times \vec n_0$. Here, we use the degenerate planar anchoring potential of Fournier and Galatola \cite{Fournier2005}   with $W_1=W_2=1.0\times10^{-2}\text{ J m}^{-2}$, which is in the strong anchoring regime. The defect structure in this case is a pair of boojums, with $+1$ winding number in the surface of the colloid, situated on opposing portions of the cylinder's two edges (Fig.~\ref{figplanar}b,c). The regions of melted nematic signalling the presence of boojums extend somewhat to follow the circular shape of the edge. Comparing to Fig.~\ref{rodsfig}c,d, we see that each boojum sits on the half of the circular edge {\em not} followed by the disclination in the case of homeotropic anchoring. 

How does our observed oblique colloidal alignment depend on the sharpness of the colloidal particle's edges? In Fig.~\ref{figsharp}, we numerically vary the sharpness parameter $p$ for cylinders modeled as supereggs, fixing the aspect ratio at 3. The equilibrium long-axis angle $\phi_0$ is plotted against $R_c/\xi_N$ in Fig.~\ref{figsharp}a, where $R_c$ is the radius of curvature of the superegg edge (the minimum radius of curvature of the superellipse cross-section) and $\xi_N\approx 3.0$ nm is the nematic correlation length. As the edge's radius of curvature increases, $\phi_0$ increases toward $90^\circ$, the value favored by ellipsoids. This increase in $\phi_0$ is slower for larger cylinders. However, the same data for $\phi_0$ collapses nicely across colloid sizes when plotted against the sharpness parameter $p$ in Fig.~\ref{figsharp}b, rather than against $R_c$. This shows that the shape, rather than absolute size scale of the edge, determines the colloid's orientation. 

The free energy compared to that of the $p=1$ ellipsoid increases as $p$ increases (decreasing $R_c$), as shown in Fig.~\ref{figsharp}c, and unsurprisingly the energy increases faster for larger supereggs. The bulk and elastic components of the free energy (respectively, integrals of the first and second lines of the integrand in Equation \ref{FLdG}) are plotted in Fig.~\ref{figsharp}d. The bulk component, chiefly the defect core energy, contributes the most to the free energy's dependence on edge sharpness. However, the contribution of the elastic energy component is also significant, and increasingly so at larger sharpness (smaller $R_c$).

Another consequence of the disclination configuration found above is that there are two energetically equivalent ground states: that of Fig.~\ref{rodsfig}c and its mirror image reflected through $\vec{n}_0$. We now wish to calculate pair potentials for colloidal cylinders in the thin planar cell geometry, which we will show to depend strongly on whether the two cylinders are in the same or mirror-image configurations.

To begin to understand how the effect of sharp edges on colloids' orientation in turn affects the properties of colloidal assembly, we seek to calculate the energy of pairs of cylindrical colloids. 
To reduce the number of variables, we fix the orientation of both cylinders in the ground-state orientation of an isolated cylinder \cite{hung2009faceted}. The remaining variables are the distance and angle of the cylinders' center-to-center separation, as well as the choice of equivalent or mirror-image ground states for the colloidal pair. 

\begin{figure}
\begin{center}
\includegraphics[width=\linewidth]{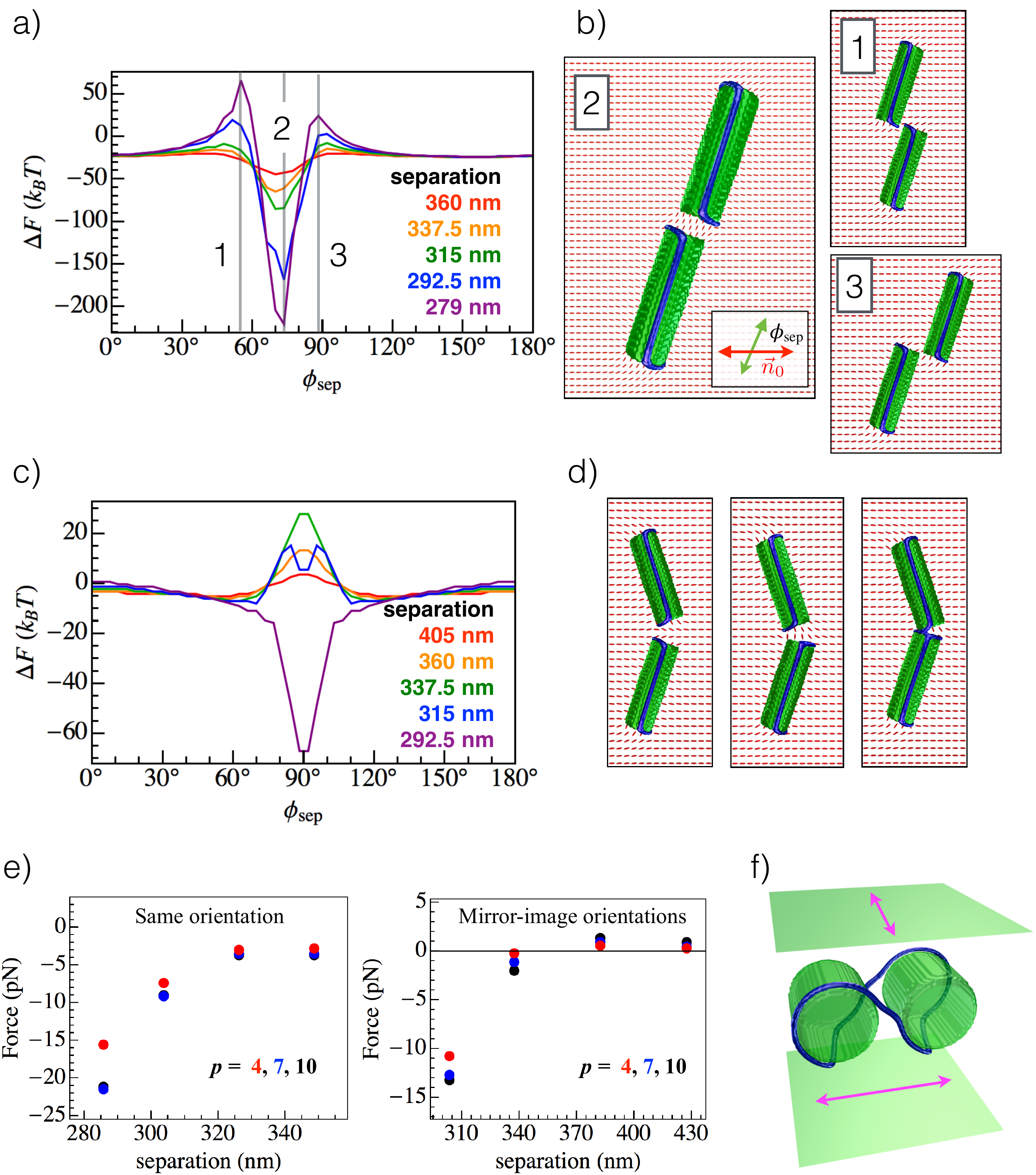} 
\caption[Pair potential for two colloidal cylinders in the same one-cylinder ground state .]{ Interaction of two colloidal cylinders. Both cylinders have diameter $2b=90$ nm, aspect ratio $a/b=3$, and sharpness $p=10$. (a) Pair potential (plus an arbitrary uniform shift) for two cylinders held at the same single-particle-equilibrium orientation, as a function of their separation angle $\phi_{\text{sep}}$ relative to $\vec n_0$, for various center-to-center separation distances (different curves). Each curve has fixed colloid center-to-center separation, as marked on the plot. (b) The cylinder separation angles with the greatest attraction or repulsion, as marked in (a). (c) Pair potential  (plus an arbitrary uniform shift) for two cylinders held in the mirror-image single-particle-equilibrium orientations. (d) Two cylinders held in the mirror-image single-particle-equilibrium orientations with $\phi_{\text{sep}}=90^\circ$, at center-to-center separations of (from left to right) 337.5, 315, and 292.5 nm. 
(e) Effect of  sharpness parameter $p$ on the effective interparticle force, at various center-to-center separations, linearly interpolated from free energy data. Left: Cylinders with the same orientation and $\phi_{\text{sep}}=\tan^{-1}(a/b)$, the attractive direction identified in (a). Right: Cylinders with mirror-image orientations and $\phi_{\text{sep}}=90^\circ$. 
(f) A topologically entangled cylinder pair in a twist cell, with magenta arrows showing the rubbing directions at the top and bottom boundaries. Here the colloid diameter is 90 nm, the aspect ratio is $a/b=1$, and $p=10$. }
\label{cylinderpairfig}
\end{center}
\end{figure}

The pair potential for two cylinders in the same ground state is shown in Fig.~\ref{cylinderpairfig}a. These cylinders attract end-to-end, as shown by the decrease in energy with decreasing center-to-center separation distance as the cylinders, oriented at angle $\phi_0$, approach along a separation angle $\phi_{\text{sep}}=\phi_0$ relative to $\vec{n}_0$, as illustrated in Fig.~\ref{cylinderpairfig}b(2). This result predicts end-to-end chaining for a series of cylinders in the equivalent ground state orientation. Whether cylinders remain oriented at an angle $\phi_0$ relative to $\vec{n}_0$ after forming a chain is an open question. LdG modeling shows that a chained cylinder pair can slightly lower its energy by fusing the two disclination rings into one, and reorienting the chained cylinders at a larger angle $\phi_0$ preferred by a single cylinder of twice the aspect ratio. 
This suggests cylinders might ``untilt'' toward $\phi=90^\circ$ as they assemble into chains. However, there may be a significant energy barrier to such a rearrangement of two disclination loops into one, just as the entangling of spherical colloids requires an input of energy, typically from laser tweezers,  to melt the nematic locally \cite{Ravnik2007}. Meanwhile, the two cylinders in Fig.~\ref{cylinderpairfig}a repel at two angles near $\phi_{\text{sep}}=\phi_0$, corresponding to the situations shown in Fig.~\ref{cylinderpairfig}b(1,3) where the approaching sides of the cylinder edges are both with or both without the disclination. 

The pair potential changes drastically when the two cylinders are instead in the different, mirror-image states (Fig.~\ref{cylinderpairfig}c). At large distances, the cylinders repel strongly when approaching along a line perpendicular to $\vec{n}_0$. However, if the cylinders are forced close enough together, one of the two disclinations switches the side of the cylinder end that it follows, replacing the ``S'' profile with a ``C'' (Fig.~\ref{cylinderpairfig}d). Thereafter, the repulsion at separation angle $\phi=90^\circ$ is replaced with an attraction that strongly binds the two cylinders together. It is possible that such a switching move in an experiment would be followed by a reorientation of the cylinder to the opposite ground-state angle and a switch of the other end's disclination kink to turn the ``C'' configuration back into an ``S'', after which the cylinders could assemble end to end. 

From the change in free energy with decreasing center-to-center separation distance at fixed $\phi_{\text{sep}}$, we can infer the inter-particle force along a specified separation direction. Fig.~\ref{cylinderpairfig}e shows these force calculations for three values of the superegg sharpness, $p=4$, $7$, and $10$. For colloids with the same orientation, we consider separation along the attractive direction $\phi_{\text{sep}}=\phi_0\approx \tan^{-1}(a/b)$ (left panel of Fig.~\ref{cylinderpairfig}e). For colloids in mirror-image configurations, we consider separation along $\phi_{\text{sep}}=90^\circ$, along which the interaction changes from repulsive to attractive with decreasing distance (right panel of Fig.~\ref{cylinderpairfig}e). We find that the sign of the force is never changed by making the edges less sharp, as the qualitative features of the free energy's dependence on the colloid separation vector do not change. However, decreasing $p$ usually reduces the magnitude of the force. This reduction varies considerably, between 0\% and 50\% for $p=7$ relative to $p=10$, and between 17\% and 87\% for $p=4$ relative to $p=10$. Of the three tested values of $p$, the $p=4$ superegg pair has the smallest-magnitude force at each value of $\phi_{\text{sep}}$. This suggests that the sharpness of the colloidal particles' edges can be manipulated to change the strength of their interactions without altering the equilibrium configuration. 

By changing the boundary conditions in the lateral directions to periodic, we can observe that cylinder chains repel one another laterally. This raises the possibility that, at high density, cylinders could self-organize into a one-dimensional lattice of tilted chains at a preferred spacing.

The fact that disclinations follow the sharp edges of cylinders also has implications for colloids' entanglement by merging of their disclinations. Such entanglement has drawn great interest for generating a variety of stable structures \cite{Ravnik2007}, including hierarchical assembly of colloids of different sizes\cite{vskarabot2008hierarchical} and a wide array of defect knots and links in colloid lattices \cite{Tkalec:2011lj}. A pair of entangled cylinders is shown in Fig.~\ref{cylinderpairfig}f, where a ``twist cell'' geometry with $90^\circ$ offset between anchoring directions at the top and bottom surfaces is used to stabilize the entanglement. As with individual cylinders, the disclination travels in a straight line roughly along the long axis of the cylinder and then follows a portion of the sharp edge at the cylinder end before traversing to the end edge of the other cylinder.

\section{Colloidal cubes}

In going from ellipsoids to cylinders and microbullets, we found that adding an edge or two created a new realignment effect with significant consequences for colloidal assembly. What will happen if the colloids are cubical, with twelve edges? Now there exist many ways for a disclination to wrap around the colloid while following {\em only} edges.

\begin{figure}
\begin{center}
\includegraphics[width=\linewidth]{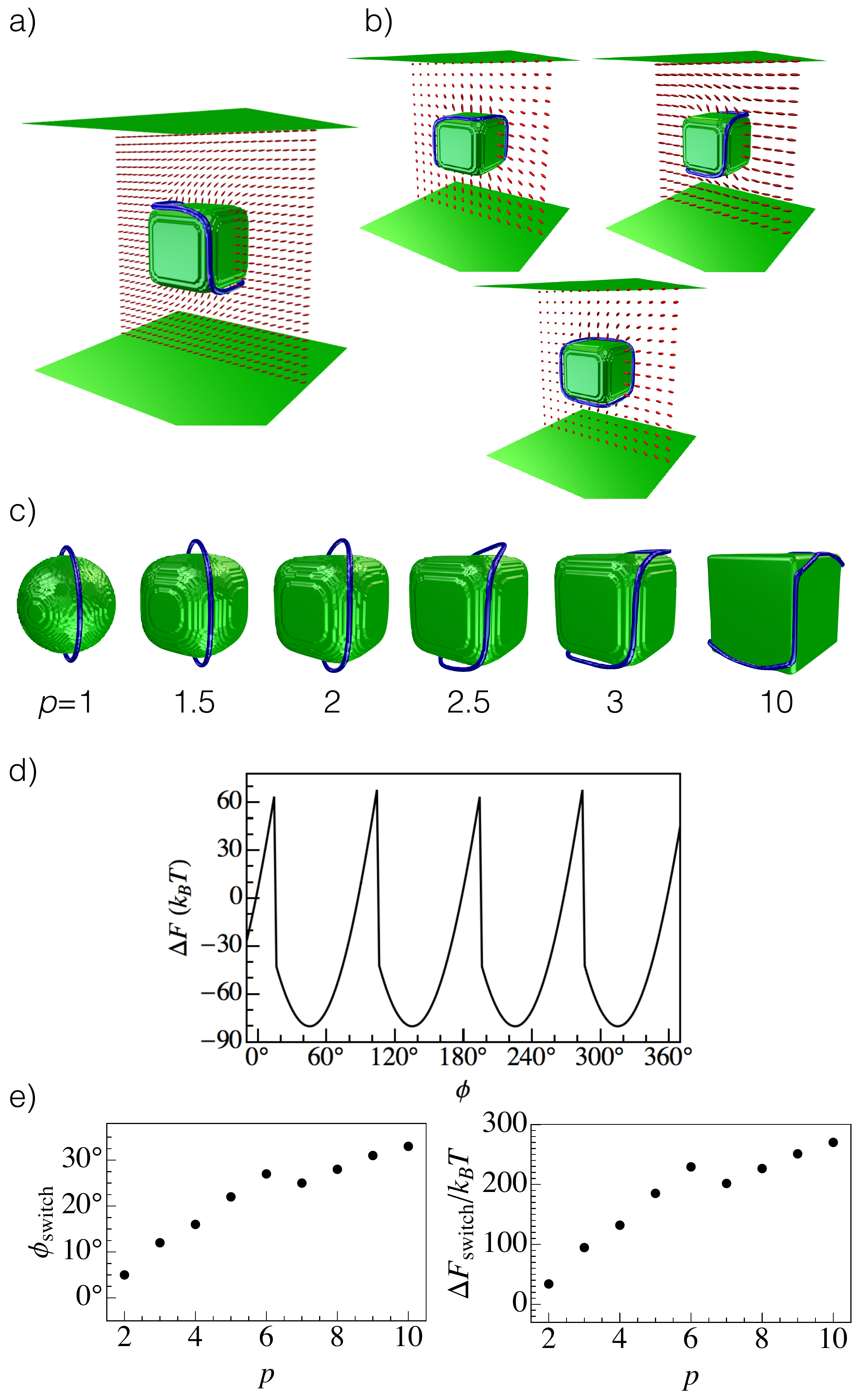}
\caption[Rotating a cubical colloid.]{Colloidal cubes.  (a) The disclination follows the edges of the cube, enclosing three faces. (b) As the cube is rotated, the disclination cycles through four possible arrangements, that shown in (a) and the three shown here. (c) Superellipsoids of diameter 540 nm at various values of sharpness $p$, showing the gradual transformation of the disclination configuration.  (d) Change in free energy as a cube is rotated about a vertical line through its center by angle $\phi$, after numerically minimizing the energy with each incremental rotation. The cube is 180 nm on each side and is modeled as a superellipsoid with sharpness $p=3$.  (e) Colloid orientation angle $\phi_{\text{switch}}$ (modulo $90^\circ$) and magnitude of the change in free energy $\Delta F_{\text{switch}}$ when the disclination configuration suddenly switches, for superellipsoids of diameter 180 nm and varying sharpness $p$.} 
\label{cubefig}
\end{center}
\end{figure}

We use LdG numerical modeling to study cubical colloids modeled using the special case of the superellipse equation, Eqn.~\ref{superellipsoideqn}, with two of the cube's faces having normals fixed in the vertical direction. 
The disclination now wraps around six of the colloid's twelve edges, dividing three faces from the other three, in one of the configurations shown in Fig.~\ref{cubefig}a,b. 
As with the rod-like colloids above, we rotate the cube about the $z$-axis to find the orientation that minimizes the free energy. We find that the cube prefers to have the normals of its side faces oriented at $45^\circ$ relative to $\vec{n}_0$. 

As the sharpness $p$ is varied, transforming a sphere continuously into an approximate cube, the disclination configuration also changes continuously, as shown in Fig.~\ref{cubefig}c. The vertical cube edges pin portions of the disclination, while above and below the cube the disclination deflects gradually toward the horizontal edges. The disclination also increasingly avoids cube vertices  as  $p$ increases, because the nematic director profile around a disclination between two cube faces is incompatible with the normal direction of a third, mutually perpendicular cube face. As a result, for large values of $p$ the disclination's otherwise straight contour is interrupted by deflections that conspicuously avoid the cube vertices, as shown in Fig.~\ref{cubefig}c for $p=10$.

Hung and Bale \cite{hung2009faceted} found numerically a similar defect configuration for smaller cubic nanoparticles of side length 40 nm in certain orientations. However, for the cube orientations considered here, Hung and Bale found that the nematic melted over a large area of two or four cube faces, whereas all defects in our findings are line-like. The discrepancy likely arises from the earlier work's smaller cube size, closer to the nematic correlation length.

The multiplicity of possible disclination arrangements has a surprising consequence: sudden reconfiguration of the disclination ring when the cube is rotated. To see this, we conduct LdG energy minimization at an initial cube orientation, then rotate the cube by small increments, re-minimizing the energy after each increment. 
This models a quasi-static rotation of the colloid relative to fixed anchoring conditions at the top and bottom surfaces. The result, shown in Fig.~\ref{cubefig}d, is that the elastic free energy builds up until a sudden rearrangement of the disclination to a set of edges more compatible with the new cube orientation, allowing an abrupt drop in free energy. 
Upon further rotation, the cube keeps the same disclination arrangement through an energy minimum with $\phi$ at odd multiples of $45^\circ$ before the elastic energy builds up again, leading to another sudden defect jump. 

Changing the sharpness $p$ of the superellipsoid used to approximate a cube quantitatively adjusts the cube orientation $\phi_{\text{switch}}$ at which this sudden switch in defect configuration occurs. Sharper edges more strongly pin the disclination, requiring a greater angle of rotation before the disclination rearrangement, and the resulting sudden drop in free energy $\Delta F_{\text{switch}}$ is of greater magnitude (Fig.~\ref{cubefig}e). Both  $\phi_{\text{switch}}$ and $\Delta F_{\text{switch}}$ vary roughly linearly with $p$.

A similar but less drastic sudden defect reconfiguration also occurs for cylindrical colloids, between the two mirror-symmetric configurations mentioned previously.


Pair potentials for colloidal cubes were also calculated. As for pairs of cylinders above, the colloid orientations were fixed in the single-cube ground state, with the faces' normals at $45^\circ$ angles to $\vec{n}_0$. The interaction that we find here is simple but interesting: Cubes attract face-to-face and repel edge-to-edge.

\section{Summary and outlook}
Through Landau-de Gennes numerical modeling of non-spherical homeotropic colloids, we have found that details of colloid shape,  in particular the presence of sharp edges, have a remarkable effect in reshaping the companion disclination ring. The disclination executes sharp turns in order to follow (portions of) the edges so as to alleviate the elastic energy cost of splay required by strong homeotropic anchoring. While this might at first seem to be only a microscopic effect, the defect's attraction to edges leads to a realignment of the colloid relative to the background director field. The colloidal realignment in turn affects the geometry of colloidal assembly, as colloids tend to attract along the directions normal to the faces with sharp edges. This predicted realignment is in qualitative agreement with our experimental results and offers an explanation for previous experimental observations of oblique orientation of colloidal micro-rods \cite{tkalec2008interactions}. We also predict sudden changes in defect arrangement as the colloid is rotated relative to the far-field director, or as colloids are forced together along a repulsive direction.

Our results suggest that it will be worthwhile to perform numerical and experimental studies on large numbers of sharp-edged colloids at high density. Self-assembled chains and lattices quite different from those formed by spheres could be produced and tuned by varying the colloids' geometrical parameters, such as aspect ratio and edge sharpness. Also, faceted colloid shapes of increased complexity could be investigated to determine the geometric possibilities offered by disclinations effectively confined to a one-dimensional subset of a three-dimensional system -- a prospect that suggests exciting connections to topology \cite{machon2013knots,vcopar2013elementary}. These areas of investigation will further the key finding presented here: Colloidal assembly and alignment in nematic liquid crystals are highly sensitive to details of colloid shape, offering a route to tunable effective colloid interactions and an expanded library of self-assembly outcomes.

%
%
%
%
%
%
%
%
%

\section*{Acknowledgments}
We thank Simon \v{C}opar, Miha Ravnik, and Randall Kamien for helpful discussions. Our numerical LdG modeling benefited greatly from the advice and assistance of Gareth Alexander and Carl Goodrich. Most visualizations of numerical results were made using a POV-Ray scene file generator provided to us by Simon \v{C}opar. This work was supported in part through the University of Pennsylvania MRSEC Grant NSF DMR11-20901. D.A.B.\ was supported by NSF Grant DGE-1321851 and by a University of Pennsylvania Teece Fellowship.


\begin{thebibliography}{10}

\bibitem{poulin1997novel}
Philippe Poulin, Holger Stark, T.C. Lubensky, and D.A. Weitz.
\newblock Novel colloidal interactions in anisotropic fluids.
\newblock {\em Science}, 275(5307):1770--1773, 1997.

\bibitem{Musevic2006}
I.~Mu{\v{s}}evi{\v{c}}, M.~{\v{S}}karabot, U.~Tkalec, M.~Ravnik, and
  S.~{\v{Z}}umer.
\newblock Two-dimensional nematic colloidal crystals self-assembled by
  topological defects.
\newblock {\em Science}, 313(5789):954--958, 2006.

\bibitem{smalyukh2004ordered}
I.I. Smalyukh, S.~Chernyshuk, B.I. Lev, A.B. Nych, U.~Ognysta, V.G. Nazarenko,
  and O.D. Lavrentovich.
\newblock Ordered droplet structures at the liquid crystal surface and
  elastic-capillary colloidal interactions.
\newblock {\em Physical Review Letters}, 93(11):117801, 2004.

\bibitem{Gharbi2011anchoring}
Mohamed~Amine Gharbi, Maurizio Nobili, Martin In, Guillaume Pr{\'e}vot, Paolo
  Galatola, Jean-Baptiste Fournier, and Christophe Blanc.
\newblock Behavior of colloidal particles at a nematic liquid crystal
  interface†.
\newblock {\em Soft Matter}, 7(4):1467--1471, 2011.

\bibitem{Skarabot2008}
M.~{\v{S}}karabot, M.~Ravnik, S.~{\v{Z}}umer, U.~Tkalec, I.~Poberaj,
  D.~Babi{\v{c}}, N.~Osterman, and I.~Mu{\v{s}}evi{\v{c}}.
\newblock Interactions of quadrupolar nematic colloids.
\newblock {\em Physical Review E}, 77(3):031705, 2008.

\bibitem{terentjev1995disclination}
E.M. Terentjev.
\newblock Disclination loops, standing alone and around solid particles, in
  nematic liquid crystals.
\newblock {\em Physical Review E}, 51(2):1330, 1995.

\bibitem{Ravnik2007}
M.~Ravnik, M.~{\v{S}}karabot, S.~{\v{Z}}umer, U.~Tkalec, I.~Poberaj,
  D.~Babi{\v{c}}, N.~Osterman, and I.~Mu{\v{s}}evi{\v{c}}.
\newblock Entangled nematic colloidal dimers and wires.
\newblock {\em Physical Review Letters}, 99(24):247801, 2007.

\bibitem{Tkalec:2011lj}
Uro{\v s} Tkalec, Miha Ravnik, Simon {\v C}opar, Slobodan {\v Z}umer, and Igor
  Mu{\v s}evi{\v c}.
\newblock Reconfigurable knots and links in chiral nematic colloids.
\newblock {\em Science}, 333(6038):62--5, 2011.

\bibitem{nych2013assembly}
A.~Nych, U.~Ognysta, M.~{\v{S}}karabot, M.~Ravnik, S.~{\v{Z}}umer, and
  I.~Mu{\v{s}}evi{\v{c}}.
\newblock Assembly and control of 3d nematic dipolar colloidal crystals.
\newblock {\em Nature Communications}, 4:1489, 2013.

\bibitem{muvsevivc2011direct}
I.~Mu{\v{s}}evi{\v{c}}, M.~{\v{S}}karabot, and M.~Humar.
\newblock Direct and inverted nematic dispersions for soft matter photonics.
\newblock {\em Journal of Physics: Condensed Matter}, 23(28):284112, 2011.

\bibitem{ravnik2011three}
Miha Ravnik, Gareth~P. Alexander, Julia~M. Yeomans, and Slobodan {\v{Z}}umer.
\newblock Three-dimensional colloidal crystals in liquid crystalline blue
  phases.
\newblock {\em Proceedings of the National Academy of Sciences},
  108(13):5188--5192, 2011.

\bibitem{damasceno2012predictive}
Pablo~F Damasceno, Michael Engel, and Sharon~C Glotzer.
\newblock Predictive self-assembly of polyhedra into complex structures.
\newblock {\em Science}, 337(6093):453--457, 2012.

\bibitem{C2SM25813G}
Ran Ni, Anjan~Prasad Gantapara, Joost de~Graaf, Rene van Roij, and Marjolein
  Dijkstra.
\newblock Phase diagram of colloidal hard superballs: from cubes via spheres to
  octahedra.
\newblock {\em Soft Matter}, 8:8826--8834, 2012.

\bibitem{de2011dense}
Joost de~Graaf, Ren{\'e} van Roij, and Marjolein Dijkstra.
\newblock Dense regular packings of irregular nonconvex particles.
\newblock {\em Physical Review Letters}, 107(15):155501, 2011.

\bibitem{botto2012capillary}
Lorenzo Botto, Eric~P Lewandowski, Marcello Cavallaro, and Kathleen~J Stebe.
\newblock Capillary interactions between anisotropic particles.
\newblock {\em Soft Matter}, 8(39):9957--9971, 2012.

\bibitem{Lee2011195}
Kyung~Jin Lee, Jaewon Yoon, and Joerg Lahann.
\newblock Recent advances with anisotropic particles.
\newblock {\em Current Opinion in Colloid \& Interface Science}, 16(3):195 --
  202, 2011.

\bibitem{tkalec2008interactions}
U.~Tkalec, M.~{\v{S}}karabot, and I.~Mu{\v{s}}evi{\v{c}}.
\newblock Interactions of micro-rods in a thin layer of a nematic liquid
  crystal.
\newblock {\em Soft Matter}, 4(12):2402--2409, 2008.

\bibitem{HungPRE2009}
Francisco~R. Hung.
\newblock Quadrupolar particles in a nematic liquid crystal: Effects of
  particle size and shape.
\newblock {\em Physical Review E}, 79:021705, 2009.

\bibitem{tasinkevych2014dispersions}
Mykola Tasinkevych, Fr{\'e}d{\'e}ric Mondiot, Olivier Mondain-Monval, and
  Jean-Christophe Loudet.
\newblock Dispersions of ellipsoidal particles in a nematic liquid crystal.
\newblock {\em Soft Matter}, 10(12):2047--2058, 2014.

\bibitem{gharbi2013microbulletPublished}
Mohamed~Amine Gharbi, Marcello Cavallaro~Jr., Gaoxiang Wu, Daniel~A. Beller,
  Randall~D. Kamien, Shu Yang, and Kathleen~J. Stebe.
\newblock Microbullet assembly: interactions of oriented dipoles in confined
  nematic liquid crystal.
\newblock {\em Liquid Crystals}, 40(12):1619--1627, 2013.

\bibitem{lapointe2009shape}
Clayton~P. Lapointe, Thomas~G. Mason, and Ivan~I. Smalyukh.
\newblock Shape-controlled colloidal interactions in nematic liquid crystals.
\newblock {\em Science}, 326(5956):1083--1086, 2009.

\bibitem{dontabhaktuni2012shape}
Jayasri Dontabhaktuni, Miha Ravnik, and Slobodan {\v{Z}}umer.
\newblock Shape-tuning the colloidal assemblies in nematic liquid crystals.
\newblock {\em Soft Matter}, 8(5):1657--1663, 2012.

\bibitem{cavallaro2013ringSoftMatter}
Marcello Cavallaro~Jr., Mohamed~A. Gharbi, Daniel~A. Beller, Simon {\v{C}}opar,
  Zheng Shi, Randall~D. Kamien, Shu Yang, Tobias Baumgart, and Kathleen~J.
  Stebe.
\newblock Ring around the colloid.
\newblock {\em Soft Matter}, 9(38):9099--9102, 2013.

\bibitem{senyuk2012topological}
B.~Senyuk, Q.~Liu, S.~He, R.D. Kamien, R.B. Kusner, T.C. Lubensky, and I.I.
  Smalyukh.
\newblock Topological colloids.
\newblock {\em Nature}, 493(7431):200--205, 2012.

\bibitem{machon2013knots}
Thomas Machon and Gareth~P. Alexander.
\newblock Knots and nonorientable surfaces in chiral nematics.
\newblock {\em Proceedings of the National Academy of Sciences},
  110(35):14174--14179, 2013.

\bibitem{cavallaro2013exploitingPNAS}
Marcello Cavallaro, Mohamed~A. Gharbi, Daniel~A. Beller, Simon {\v{C}}opar,
  Zheng Shi, Tobias Baumgart, Shu Yang, Randall~D. Kamien, and Kathleen~J.
  Stebe.
\newblock Exploiting imperfections in the bulk to direct assembly of surface
  colloids.
\newblock {\em Proceedings of the National Academy of Sciences},
  110(47):18804--18808, 2013.

\bibitem{PhysRevLett.112.225501}
Nuno~M. Silvestre, Qingkun Liu, Bohdan Senyuk, Ivan~I. Smalyukh, and Mykola
  Tasinkevych.
\newblock Towards template-assisted assembly of nematic colloids.
\newblock {\em Physical Review Letters}, 112:225501, 2014.

\bibitem{2014arXiv1406.0702E}
Z.~{Eskandari}, N.~M. {Silvestre}, M.~M. {Telo da Gama}, and M.~R. {Ejtehadi}.
\newblock {Particle selection through topographic surface patterns in nematic
  colloids}.
\newblock {\em arXiv preprint}, arXiv:1406.0702, 2014.

\bibitem{tsakonas2007multistable}
C.~Tsakonas, A.J. Davidson, C.V. Brown, and N.J. Mottram.
\newblock Multistable alignment states in nematic liquid crystal filled wells.
\newblock {\em Applied Physics Letters}, 90(11):111913--111913, 2007.

\bibitem{luo2012multistability}
Chong Luo, Apala Majumdar, and Radek Erban.
\newblock Multistability in planar liquid crystal wells.
\newblock {\em Physical Review E}, 85(6):061702, 2012.

\bibitem{conradi2009janus}
Marjetka Conradi, Miha Ravnik, Marjan Bele, Milena Zorko, Slobodan {\v{Z}}umer,
  and Igor Mu{\v{s}}evi{\v{c}}.
\newblock Janus nematic colloids.
\newblock {\em Soft Matter}, 5(20):3905--3912, 2009.

\bibitem{vcopar2014janus}
Simon {\v{C}}opar, Miha Ravnik, and Slobodan {\v{Z}}umer.
\newblock Janus nematic colloids with designable valance.
\newblock {\em Materials}, 7(6):4272--4281, 2014.

\bibitem{stark1999director}
Holger Stark.
\newblock Director field configurations around a spherical particle in a
  nematic liquid crystal.
\newblock {\em The European Physical Journal B-Condensed Matter and Complex
  Systems}, 10(2):311--321, 1999.

\bibitem{grollau2003spherical}
Sylvain Grollau, N.L. Abbott, and J.J. de~Pablo.
\newblock Spherical particle immersed in a nematic liquid crystal: Effects of
  confinement on the director field configurations.
\newblock {\em Physical Review E}, 67(1):011702, 2003.

\bibitem{Ravnik2009a}
Miha Ravnik and Slobodan \v{Z}umer.
\newblock Landau-de gennes modelling of nematic liquid crystal colloids.
\newblock {\em Liquid Crystals}, 36(10-11):1201--1214, 2009.

\bibitem{gharbi2013microparticles}
Mohamed~Amine Gharbi, David Se{\v{c}}, Teresa Lopez-Leon, Maurizio Nobili, Miha
  Ravnik, Slobodan {\v{Z}}umer, and Christophe Blanc.
\newblock Microparticles confined to a nematic liquid crystal shell.
\newblock {\em Soft Matter}, 9(29):6911--6920, 2013.

\bibitem{hung2009faceted}
F.R. Hung and S.~Bale.
\newblock Faceted nanoparticles in a nematic liquid crystal: defect structures
  and potentials of mean force.
\newblock {\em Molecular Simulation}, 35(10-11):822--834, 2009.

\bibitem{mermin1979topological}
N.~David Mermin.
\newblock The topological theory of defects in ordered media.
\newblock {\em Reviews of Modern Physics}, 51(3):591, 1979.

\bibitem{Lubensky1998}
T.C. Lubensky, D.~Pettey, N.~Currier, and H.~Stark.
\newblock Topological defects and interactions in nematic emulsions.
\newblock {\em Physical Review E}, 57(1):610, 1998.

\bibitem{Fournier2005}
J.-B. Fournier and P.~Galatola.
\newblock Modeling planar degenerate wetting and anchoring in nematic liquid
  crystals.
\newblock {\em EPL (Europhysics Letters)}, 72(3):403, 2005.

\bibitem{vskarabot2008hierarchical}
Miha {\v{S}}karabot, Miha Ravnik, Slobodan {\v{Z}}umer, Uro{\v{s}} Tkalec, Igor
  Poberaj, Du{\v{s}}an Babi{\v{c}}, and Igor Mu{\v{s}}evi{\v{c}}.
\newblock Hierarchical self-assembly of nematic colloidal superstructures.
\newblock {\em Physical Review E}, 77(6):061706, 2008.

\bibitem{vcopar2013elementary}
Simon {\v{C}}opar, Noel~A Clark, Miha Ravnik, and Slobodan {\v{Z}}umer.
\newblock Elementary building blocks of nematic disclination networks in
  densely packed 3d colloidal lattices.
\newblock {\em Soft Matter}, 9(34):8203--8209, 2013.

\end{thebibliography}

\end{document}